\documentclass[twocolumn]{article} 
\usepackage{geometry}
\usepackage{graphicx} 
\usepackage{caption}  
\usepackage{amsmath}
\title{Hidden Echoes Survive Training in Audio To Audio Generative Instrument Models}

\usepackage{bigfoot}

\DeclareNewFootnote{AAffil}[arabic]
\DeclareNewFootnote{ANote}[fnsymbol]

\usepackage{etoolbox}
\makeatletter
\patchcmd\maketitle{\def\@makefnmark{\rlap{\@textsuperscript{\normalfont\@thefnmark}}}}{}{}{}
\makeatother

\makeatletter
\def\thanksAAffil#1{
  \footnotemarkAAffil\protected@xdef\@thanks{\@thanks%
        \protect\footnotetextAAffil[\the \c@footnoteAAffil]{#1}}
}
\def\thanksANote#1{%
  \footnotemarkANote%
  \protected@xdef\@thanks{\@thanks%
        \protect\footnotetextANote[\the \c@footnoteANote]{#1}}
}
\makeatother

\author{
  Christopher J. Tralie\thanksAAffil{Ursinus College Mathematics, Computer Science, And Statistics}
  , %
  Matt Amery%
  \thanksAAffil{Create And Innovate UK}
  , %
  Benjamin Douglas\footnotemarkAAffil[1]
  , %
  Ian Utz\footnotemarkAAffil[1]
}

\begin{document}
\date{}

\maketitle

\begin{abstract}
As generative techniques pervade the audio domain, there has been increasing interest in tracing back through these complicated models to understand how they draw on their training data to synthesize new examples, both to ensure that they use properly licensed data and also to elucidate their black box behavior. In this paper, we show that if imperceptible echoes are hidden in the training data, a wide variety of audio to audio architectures (differentiable digital signal processing (DDSP), Realtime Audio Variational autoEncoder (RAVE), and ``Dance Diffusion'') will reproduce these echoes in their outputs. Hiding a single echo is particularly robust across all architectures, but we also show promising results hiding longer time spread echo patterns for an increased information capacity. We conclude by showing that echoes make their way into fine tuned models, that they survive mixing/demixing, and that they survive pitch shift augmentation during training. Hence, this simple, classical idea in watermarking shows significant promise for tagging generative audio models. 

\end{abstract}

\section{Introduction}

We seek to understand how generative audio neural network models use their training data, both to detect training on unlicensed data and to understand the inner workings of models.  One post-hoc approach is to correlate synthesized outputs from the models with specific sounds that could be in the training data \cite{batlle2024towards, barnett2024exploring}.  Other approaches modify the generator directly to watermark its outputs, such as \cite{cao2023invisible} who were inspired by \cite{wen2023tree} in the image domain.  In our work, on the other hand, we assume the least knowledge/control over the models that are used and instead restrict our focus to techniques that sit the earliest in the pipeline: {\em those that modify the training data only}. One such line of work seeks to watermark training data in such a way that when models are fine tuned, they will fail to reproduce the training data.  These so-called ``poisoning'' techniques are popular in the image processing domain (e.g. ``Glaze'' \cite{shan2023glaze} and ``Nightshade'' \cite{shan2023prompt}), and similar works have begun to appear in singing voice cloning \cite{chen2024proactive} and music generation \cite{barnett2024defenses, meerza2025harmonycloak}.  In our work, though, we do not seek to influence the behavior of the model so drastically, but rather to ``tag'' the data in such a way that the model reproduces the tag, similarly to how \cite{ditria2023hey} watermark their training data for a diffusion image model.  We are also inspired by the recent lawsuit by Getty Images against Stable Diffusion when it was discovered that the latter would often reproduce the former's watermarks in its output \cite{vincent2023getty}.  We would like to do something similar with audio, but to keep it imperceptible.

All of the above approaches use neural networks to create watermarks for generative models, but we are unaware of any works that use any simpler classical, hand-crafted audio watermarks for this purpose.  If such watermarks could survive training, this could make it simpler for practitioners to implement, and it may also more easily shed light on the inner workings of the generative models.  While many options are available, such as spread spectrum \cite{kirovski2001robust}, phase-based \cite{xiaoxiao_dong_data_2004, malik_robust_2007}, and OFDM \cite{eichelberger_receiving_2019}, we surprisingly find success with some of the oldest and simplest techniques based on echo hiding \cite{gruhl1996echo} and followup work on time-spread echo hiding \cite{ko2005time}.  If we embed a single echo or a fixed pseudorandom time-spread echo pattern across each clip in the training data, the pattern will be recreated by a variety of architectures when synthesizing new sounds.  To show this in a general, reproducible way, we test it using three open architectures with fundamentally different approaches whose code is readily available online: RAVE \cite{caillon2021rave}\footnote{ \url{https://github.com/acids-ircam/RAVE} } Dance Diffusion \cite{evans2022dancediffusion} \footnote{ \url{https://github.com/harmonai-org/sample-generator} }, and differentiable digital signal processing (DDSP) \cite{engelddsp}\footnote{We use our own vanilla implementation of DDSP at \url{https://github.com/ctralie/ddsp}}.  Each model is trained on audio only, as opposed to those also involving language models (e.g. \cite{evans2024fast}) or MIDI (e.g. \cite{hawthornemulti}), and each model is trained on a collection of instrument sounds from the same instrument.  Specifically, we train models with different conditions on each of three open datasets to further enhance reproducibility: Groove \cite{groove2019}, VocalSet \cite{wilkins2018vocalset}, and GuitarSet \cite{xi2018guitarset}, which span vocals, and drums, and acoustic guitar, respectively.  We evaluate each model using the respective vocals, drums, and ``other'' stems in the MUSDB18-HQ dataset \cite{musdb18-hq} as inputs to models trained under various conditions.  

\section{Methods}

Below we describe the generative audio to audio models we use, as well as the scheme we use to watermark the training data. Every audio sample in the training sets is converted to a 44100hz mono, as are all of the inputs to the models.  Supplementary audio examples and source code can be found at \url{https://www.ctralie.com/echoes}.

\subsection{Audio To Audio Models}

We restrict the focus of our work to audio to audio models, in which a neural network is trained on a corpus and it synthesizes outputs in the style of the corpus.  For instance, one could train such a model on a corpus of violins and feed it singing voice audio to create a ``singing violin.''  The first such technique we use, Differentiable Digital Signal Processing (DDSP) \cite{engelddsp} has the simplest architecture out of all of the models.  We use the version from the original paper in which the encoder is fixed as a 2 dimensional representation of pitch and loudness, respectively.  These dimension are then fed to a decoder network which learns an additive and subtractive synthesizer to best match the training data for a particular pitch/loudness trajectory.  The only thing we change is that we use the more recent PESTO \cite{riou2023pesto} instead of CREPE \cite{kim2018crepe} for efficiency, and we use a 3D latent space of loudness, pitch, and {\em pitch confidence}.  In the end, our DDSP models have $\approx$5 million parameters.

The second most complex model we use is ``RAVE'' \cite{caillon2021rave}, which is a two-stage model that first learns a general audio autoencoder and then improves this autoencoder with generative adversarial training.  We use Rave V2, which has $\approx$32 million parameters, and we use snake activations and train with compression augmentation. 

The most complex model we use is ``Dance Diffusion,'' which uses a vanilla diffusion network \cite{sohl2015deep} with attention to progressively denoise outputs from a completely random input.  To condition a style transfer to sound more like a particular input $x$, one can jump-start the diffusion process with a scaled $x$ and some added AWGN noise with standard deviation $\eta \in [0, 1]$.  The closer $\eta$ is to 1, the more the output will take on the character of the corpus on which Dance Diffusion was trained.  We use $\eta=0.2$ in all of our experiments, and we use a 81920 sample size, which means the receptive field spans $\approx$1.86 seconds, and the denoising network has $\approx$222 million parameters.

\subsection{Echo Hiding}
\label{sec:echohiding}

\begin{figure}
    \centering
    \includegraphics[width=\columnwidth]{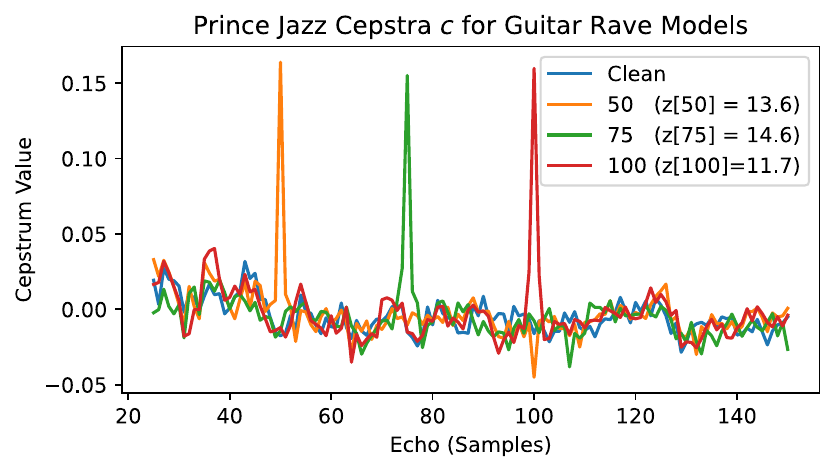}
    \caption{An example of cepstra computed on style transfer of a 30 second excerpt of a Prince jazz session at Loring Park.  RAVE models trained on data with different echoes at 50, 75, and 100 lead to visible peaks at the respective places in their ceptra on the synthesized clips.}
    \label{fig:ravecepstra}
\end{figure}

Given a discrete audio ``carrier waveform'' $x$, audio watermarking techniques hide a binary payload in a watermarked waveform $\hat{x}$ so that $x$ and $\hat{x}$ are perceptually indistinguishable.  The original echo hiding paper by \cite{gruhl1996echo} accomplishes this by creating two waveforms $x_0$ and $x_1$, each with a single echo; 
\begin{equation}
    \label{eq:echohidesingle}
    \begin{aligned}
        x_0[n] &= x[n] + \alpha x[n - \delta_0] \\
        x_1[n] &= x[n] + \alpha x[n - \delta_1]
    \end{aligned}
\end{equation}

 where $\alpha < 1$ trades off perceptibility and robustness of the watermark, and $\delta_0, \delta_1 \leq $ 100 samples at a 44.1khz sample rate.  These waveforms are then mixed together in windows to create $\hat{x}$ according to the payload; where $x_0$ is fully mixed at the center of a window if the payload contains a 0 at that moment and $x_1$ is fully mixed in if the payload contains a 0.  For a window of 1024 samples, for instance, this amounts to $\approx$43 bits per second at 44.1khz.  Because the echoes are at such a small shift, temporal aliasing of human hearing makes them less noticeable.  Furthermore, since convolution in the time domain is multiplication in the frequency domain, the logarithm of the magnitude of the DFT of a window additively separates the frequency response of the echo from the frequency response of $x$.  Therefore, the so-called ``cepstrum'' of a windowed signal $x_w$:
 
\begin{equation}
    \label{eq:cepstrum}
    c = \text{ifft} ( \log ( | \text{fft} (x_w) | ) )
\end{equation}
 
yields a signal in which a single echo is a high peak, which is referred to as the ``cepstrum'' $c$ \footnote{\cite{gruhl1996echo} note that it is more mathematically correct take the {\em complex logarithm} of the DFT before taking the inverse DFT, and they further enhance with an autocorrelation.  But we found better results with the traditional cepstrum.}.  Thus, to decode the payload from the watermarked signal, one computes $c$ on each window and infers a 0 if $c[\delta_0] > c[\delta_1]$ or a 1 otherwise.

Since we seek to hide echoes in the training data for generative models, it is unlikely that the models we train will synthesize the windows in the same order they occur in the training set.  Therefore, we do away with the windowing completely and instead hide {\em the same echo} $\delta$ in {\em the entire audio clip} of {\em each waveform in the training data}.  We then examine the cepstrum $c$ of an entire clip that comes out of our models.  To score the cepstrum value at $\delta$ in a loudness-independent way, we compute the {\em z-score} at each lag $i$ as follows.  First, let $\mu_{c}^{a,b}[i]$ be the mean of $c$ on the interval $[a, b]$, excluding $i$:

\begin{equation}
    \mu_{c}^{a,b}[i] = \left( \sum_{\substack{j=a \\ j \neq i}}^{b} c[j] \right) / (b-a)
\end{equation}

and let $\sigma_{c}^{a,b}[i]$ be the analogous standard deviation:

\begin{equation}
    \sigma_{c}^{a,b}[i] = \sqrt{ \left( \sum_{\substack{j=a \\ j \neq i}}^{b} (c[j] - \mu_c^{a,b}[i])^2 \right) / (b-a)}
\end{equation}

then we define the z-score as:

\begin{equation}
    \label{eq:zscore}
    z_{c}^{a,b}[i] = \mu_{c}^{a,b}[i] / \sigma_{c}^{a,b}[i]
\end{equation}

A model trained on data watermarked with echo $\delta$ works well if $z^{a,b}_{\delta}[\delta] > z^{a,b}_{i}[i], i \neq \delta$.  In our experiments, we use $\alpha = 0.4$, $\delta \in \{50, 76, 76, 100\}$, $a=25$, and $b=125$.  Henceforth, we will assume those parameters and simply refer to these numbers as ``the z-scores $z$.''  Figure~\ref{fig:ravecepstra} shows example cepstra from clips created with different RAVE\cite{caillon2021rave} models trained on the GuitarSet \cite{xi2018guitarset} dataset, with various echoes $\delta$.  The peaks and z-scores show that the models reproduce the echoes they were trained on.  We will evaluate this more extensively in the experiment section (Figure~\ref{fig:singleechotable}).

\subsection{Time Spread Echo Patterns}

\begin{figure}
    \centering
    \includegraphics[width=\columnwidth]{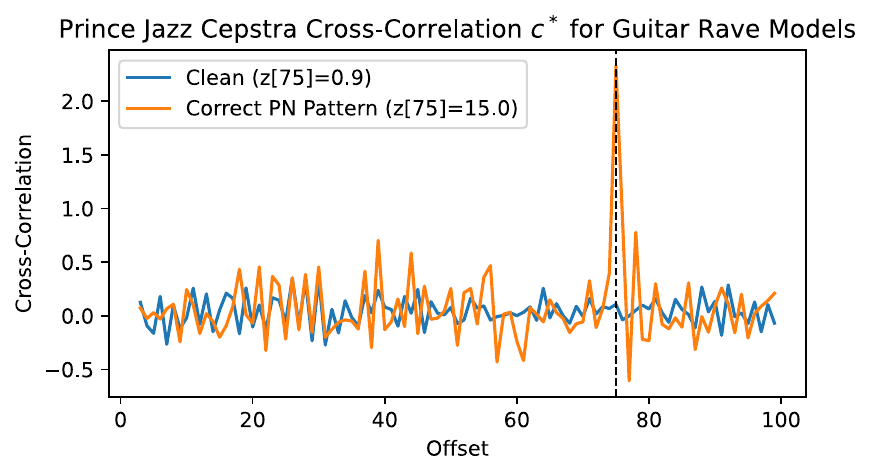}
    \caption{Comparing a 30 second style transfer using a RAVE model with a time spread echo pattern $p$ embedded in the training data to one without any pattern.  The cross-correlation of the cepstrum with $p$ peaks for the model with the embedded pattern. }
    \label{fig:ravepncepstra}
\end{figure}

\begin{figure}
    \centering
    \includegraphics[width=\columnwidth]{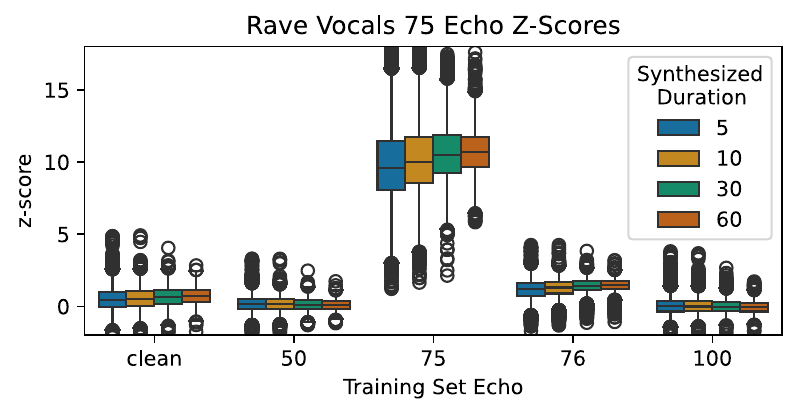}
    \caption{As this example with various tagged VocalSet training data shows, the z-scores for a 75 echo are much higher for the models that are trained on a dataset with a 75 echo embedded in every clip, and the separation increases with increasing clip duration.}
    \label{fig:ravezscoreexamples}
\end{figure}

Though we have found single echoes to be robust, the information capacity is low.  Supposing we use echoes between 50 and 100 at integer values, we can store at most $\approx$5.7 bits of information in a single dataset.  To increase the information capacity, we also explore followup work on ``time-spread echo hiding'' \cite{ko2005time} that hides an entire pseudorandom binary sequence $p$ with $L$ bits by scaling, time shifting, and convolving it with the carrier signal $x$:

\begin{equation}
\hat{x} = x * \alpha p_{\delta}, \text{ where } p_{\delta}[n] = 2 p[n - \delta] - 1
\end{equation}

where, to maintain perceptual transparency, $\alpha$ is generally significantly smaller than it is for a single echo; we use $\alpha = 0.01$.  To uncover the hidden pattern, one computes the cepstrum $c$ according to Equation~\ref{eq:cepstrum}, and then does a cross-correlation of $c$ with $(2p - 1)$ to obtain a signal $c^*$.  If the echo pattern is well preserved, then $c^*[\delta] > c^*[i \neq \delta]$.

As in the original echo hiding paper, this work hides $p$ at different offsets $\delta$ in two different signals for hiding a 1 or a 0, but, once again, we hide the same time spread echo at the same lag $\delta=75$ for the entire clip in the training data of our models.  We then compute the $z$-score $z_{c^*}^{a,b}$ on $c^*$ on the model outputs using an equation analogous to Equation~\ref{eq:zscore}, though we set $a = 3, b=L+\delta$, and we also exclude the samples of $c^*$ 3 to the left and 3 to the right when computing $\mu_{c^*}^{a,b}$ and $\sigma_{c^*}^{a,b}$.  Overall, we create 8 different versions of each training set we have, each embedded with a different time spread echo pattern of length $L=1024$.  Furthermore, we ensure that the 28 pairwise Hamming distances between the 8 time spread patterns are approximately uniformly distributed between 0 and 1024.  Figure~\ref{fig:ravepncepstra} shows an example of a style transfer on a model trained on data with the first time spread pattern embedded in all of the training data.

Note that followup work by \cite{xiang2010effective} suggests ensuring that the time spread echo patterns don't have more than two 0's or two 1's in a row, which skews the perturbations in $\hat{x}$ to less perceptible higher frequencies.  In this case, one can also compute an enhanced cross-correlation signal as $c^*[n] - 0.5c^*[n-1] - 0.5c^*[n+1]$.  Though we ensured that our time spread echo patterns satisfied this property, we did not find an improvement in our experiments, so we stick to the original cross-correlation z-score.

\section{Experiments}

\begin{figure*}
    \centering
    \includegraphics[width=\textwidth]{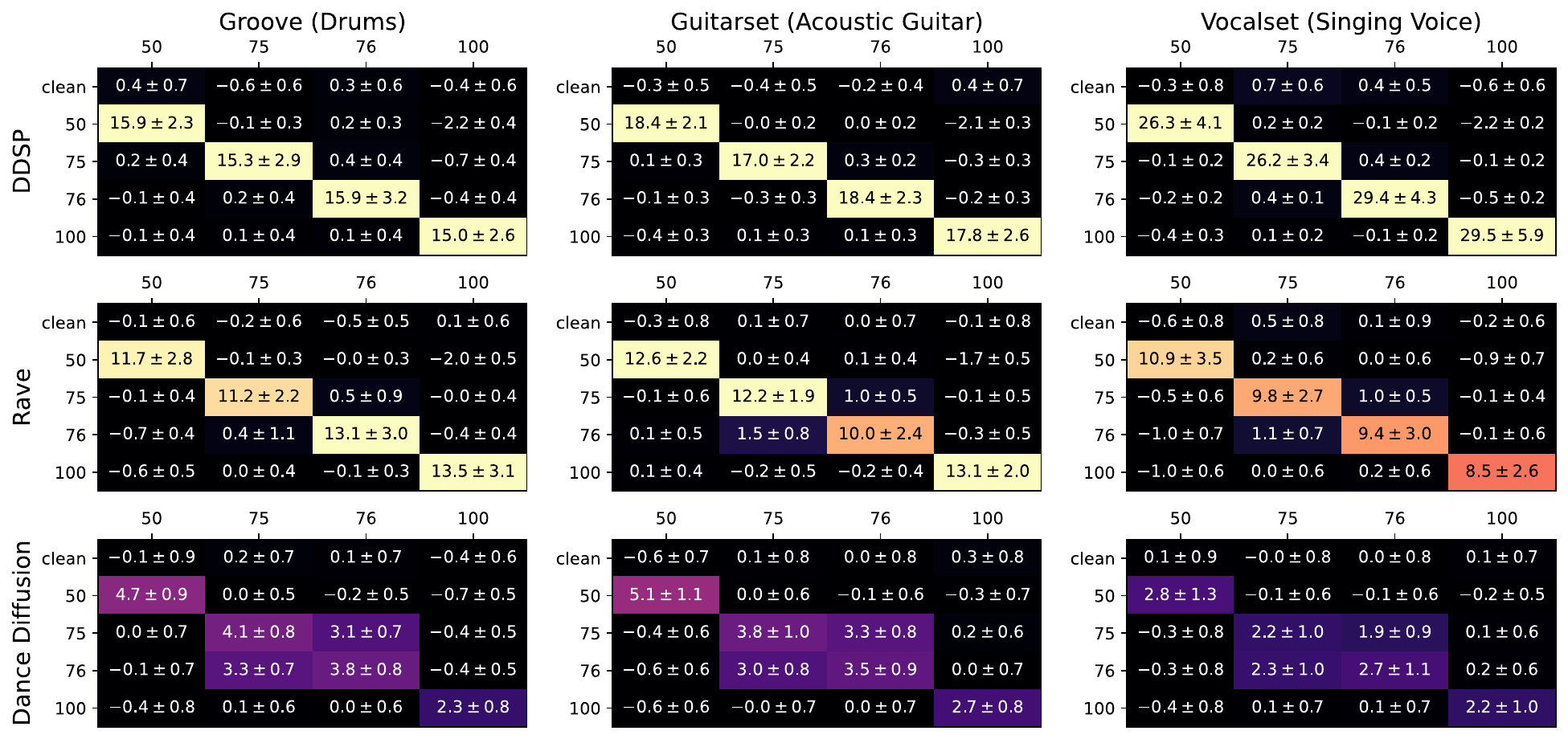}
    
    \caption{The means and standard deviations of z-scores for datasets embedded with various single echoes (along each inner row) evaluated for different echoes (along each inner column) show that all architectures (outer rows) only strongly reproduce the echoes that they were trained on across all datasets (outer columns).}
    \label{fig:singleechotable}
\end{figure*}

\begin{figure}
    \centering
    \includegraphics[width=\columnwidth]{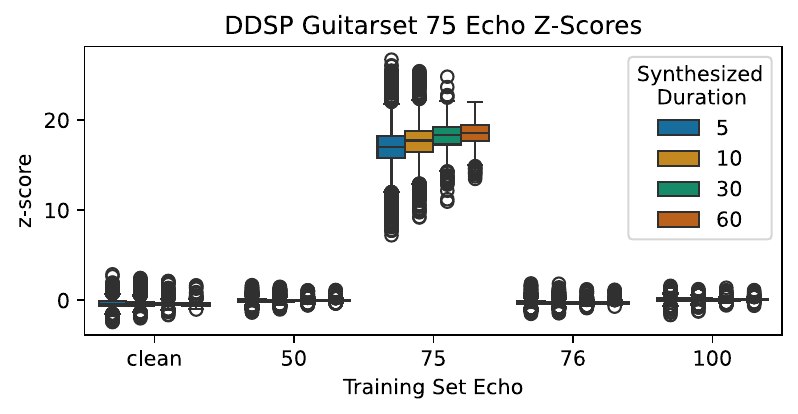}
    \caption{DDSP models show the strongest preservation of echoes over all model types, as measured by the z-score.}
    \label{fig:ddspzscoreexamples}
\end{figure}

To rigorously evaluate the efficacy of our echo watermarks, we train each of our three different model architectures on 3 different datasets:  the training set for Groove \cite{groove2019} ($\approx$8 hours), the entire VocalSet dataset \cite{wilkins2018vocalset} ($\approx$6 hours), and the entire GuitarSet dataset \cite{xi2018guitarset} ($\approx$3 hours).  For each model+architecture combination, we train a variety of models with different embedded echo patterns in the training set.  Once each model is trained, we send through as input multiple random segments of lengths 5, 10, 30, and 60 seconds, drawn from each of the 100 corresponding stems in the MUSDB18-HQ dataset \cite{musdb18-hq}.  In particular, models trained on VocalSet get the ``vocals'' stems, models trained on Groove get the ``drums'' stems, and models trained on Guitarset get ``other'' stems (which are mostly acoustic and electric guitar).  Finally, we report z-scores for various single echo and time spread echo patterns on the outputs of the models.

We train RAVE for 1.3 million steps for Groove and 2 million steps for GuitarSet and VocalSet.  We train Dance Diffusion for 50,000 steps on all models, and we train DDSP for 500,000 samples on all models.  

\subsection{Single Echo Experiments}
\label{sec:experimentssingleecho}

\begin{figure}
    \centering
    \includegraphics[width=\columnwidth]{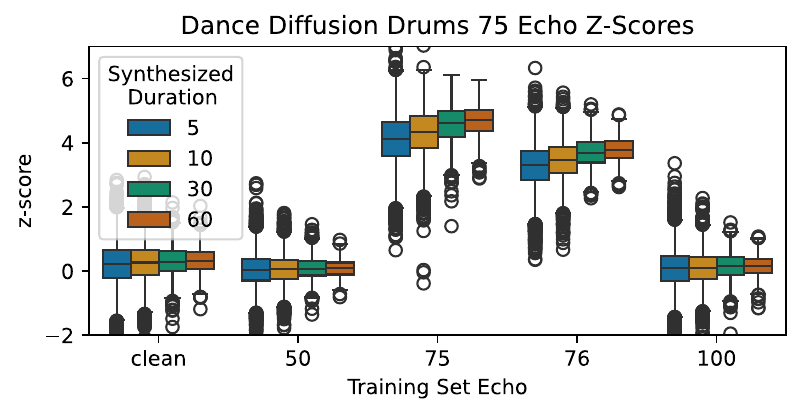}
    \caption{Dance Diffusion models show slightly weaker z-scores that may be mixed up between adjacent echoes, but they still reproduce the correct echoes overall.}
    \label{fig:dancediffusionzscoreexamples}
\end{figure}

For these experiments, we train each architecture on each of the original VocalSet, GuitarSet, and Groove datasets, as well as on each of these datasets with an embedded echo of 50, 75, 76, and 100.  Figures~\ref{fig:ravezscoreexamples},~\ref{fig:ddspzscoreexamples}, and~\ref{fig:dancediffusionzscoreexamples} show distributions of z-scores for models trained with an echo of 75 and tested with the corresponding stems.  Figure~\ref{fig:singleechotable} shows the mean and standard deviation of z-scores for the MUSDB18-HQ clips over all architectures over all instruments over all echoes.  The echoes are quite robust over all architectures.  The only weakness is a mixup of the adjacent echoes 75 and 76 for the Dance Diffusion models.


\subsection{Time Spread Echo Sequences}

\begin{figure}
    \centering
    \includegraphics[width=\columnwidth]{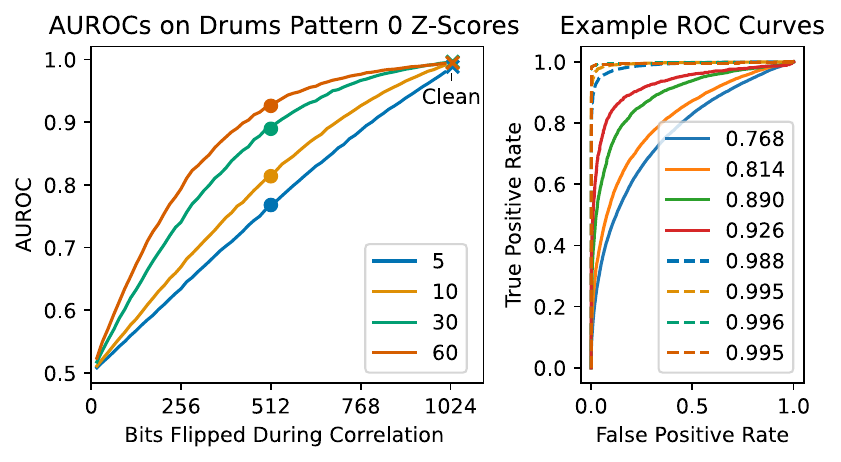}
    \caption{An example of our evaluation for a Rave model embedded with a time-spread echo pattern.  The further away the perturbed correlated pattern $p'$ gets from the truly embedded pattern $p$, the smaller the z-scores get, increasing the AUROC of z-scores from $p$ and $p'$.  Increasing the length of the synthesized clip (depicted as color) also leads to stronger detection capability.  Finally, the z-scores of $p$ in the embedded model are easily distinguishable from the z-scores of $p$ in the clean model (x's in left plot, dotted lines in right plot).}
    \label{fig:drumspn7}
\end{figure}

Next, we train RAVE and DDSP on 8 time spread echo patterns embedded in each dataset.  We omit fully training dance diffusion with these patterns due to computational constraints and poorer results.  Once again, we compute z-scores on the outputs of multiple random clips from the 100 examples in the MUSDB18-HQ training set.  To quantify the extent to which each model captures the time spread pattern, we compute z-scores on the $c^*$ correlating with the original pattern $p$ on the output cepstra, and we also compute z-scores after correlating with a perturbed version $p'$ of $p$ with an increasing number of bits randomly flipped.  To quantify how the z-scores change, we compute an ROC curve, where the true positives are z-scores correlating to $p$, and the false positives are correlating to the perturbed versions $p'$.

\begin{figure}
    \centering
    \includegraphics[width=\columnwidth]{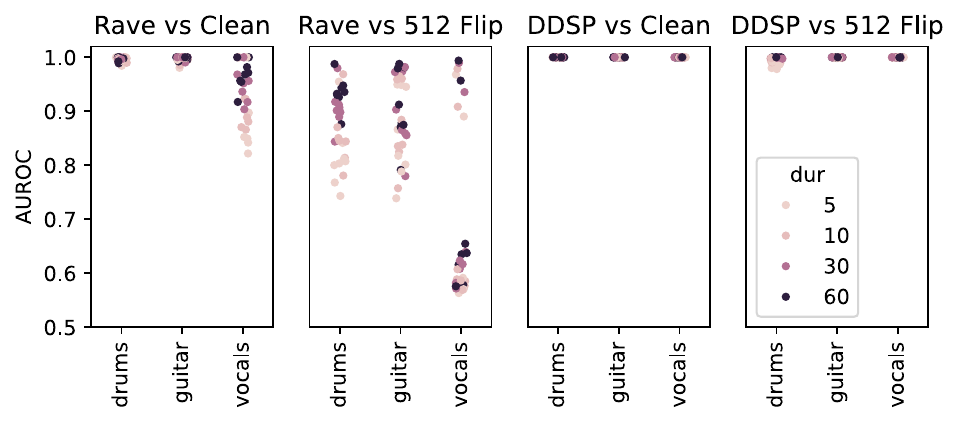}
    \caption{Z-scores of longer clips from models trained on time-spread echo patterns stand out more.}
    \label{fig:pneval}
\end{figure}

Figure~\ref{fig:drumspn7} shows an example of this evaluation on a Rave model trained on the first time spread echo pattern embedded in the Groove dataset.  The right plot shows the corresponding ROC curves for 512 bits flipped at different durations, as well as ROC curves where the false positives are z-scores in a clean model correlating $p$.  Figure~\ref{fig:pneval} shows the AUROC for all 8 pseudorandom patterns when comparing to 512 random bits flipped and when comparing to the clean model.  Inter-model comparisons of z-scores are more challenging for the Rave models compared to single echoes due to the variation in embedding strength from model to model.   However, within each model we always get a positive slope in AUROC vs bits flipped, and we can always tell the difference with the clean model.  This indicates that the correct echo patterns survive training.

\section{Additional Use Cases}

\subsection{Dance Diffusion Fine Tuning}

\begin{figure}
    \centering
    \includegraphics[width=\columnwidth]{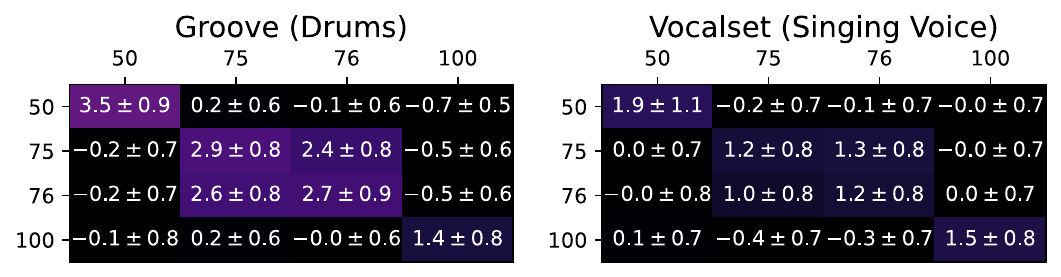}
    \caption{Fine tuning clean dance diffusion models on single echoes embeds echoes somewhat}
    \label{fig:finetuning}
\end{figure}

We use the train/test/validation set from Groove, and we create our own train/test/validation set for VocalSet (we omit GuitarSet in this experiment because it's too small).  We then embed echoes in the test set and fine tune the corresponding Dance Diffusion trained on the clean training sets, using the validation set to make sure we're not overfitting.  This represents a more realistic scenario than training such a large model from scratch with the same echo in the entire dataset.  Figure~\ref{fig:finetuning} shows the results, which show some initial promise.

\subsection{Single Echo Demixing}

\begin{figure*}
    \centering
    \includegraphics[width=\textwidth]{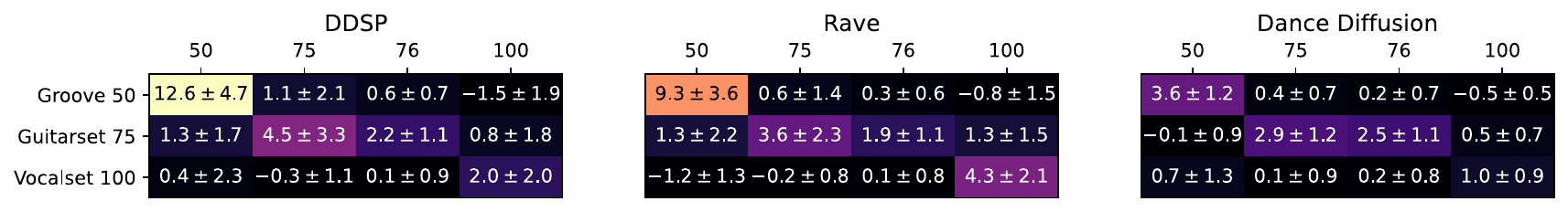}
    \caption{If we first mix together outputs of models trained on Groove with an echo of 50, GuitarSet with an echo of 75, and VocalSet with an echo of 100, the correct echoes pop out in the demixed tracks.}
    \label{fig:demucszscores}
\end{figure*}

In realistic applications of audio to audio style transfer, it is common to treat the result as a stem and {\em mix} it in with other tracks.  Hence, we perform a cursory experiment to see the extent to which the synthesized echoes survive mixing and demixing.  We use the ``hybrid Demucs'' algorithm \cite{defossez2019music} to demix the audio.  This demixing model was trained on (among other data) the MUSDB18-HQ training set, so we switch the inputs to the 50 clips from the MUSDB18-HQ {\em test set}.  

To create our testing data, for each architecture, we input the drums stem to the model trained on Groove with a 50 sample echo, the ``other'' stem to the model trained on the GuitarSet data with a 75 sample echo, and the vocals stem to the model trained on VocalSet with a 100 echo.  We then mix the results together with equal weights and demix them with Demucs into the drums, vocals, and ``other'' track.  Finally, we compute z-scores on each demixed track at echoes of 50, 75, 76, and 100.  Figure~\ref{fig:demucszscores} shows the results.  The trends are similar to the overall single echo z-scores in in Figure~\ref{fig:singleechotable}, albeit with slightly weaker z-scores.  Still, all of the correct echoes pop out in their corresponding tracks.

\subsection{RAVE Pitch Shift Augmentation}

\begin{figure}
    \centering
    \includegraphics[width=\columnwidth]{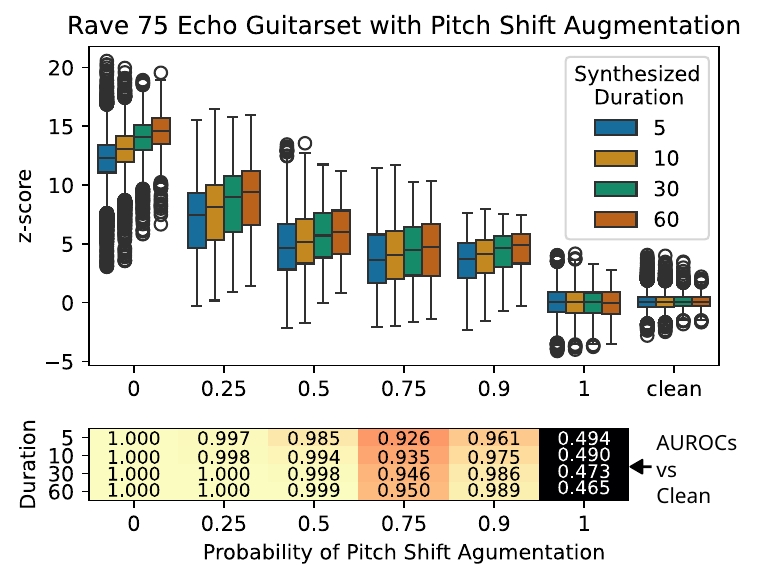}
    \caption{Z-scores generally decrease for an increasing probability of pitch augmentation, though they remain detectable even for high rates of augmentation.}
    \label{fig:pitchshiftaugmentation}
\end{figure}

Data augmentation is often important to train generalizable models.  One form of data augmentation commonly used in audio is pitch shifting.  Unfortunately, classical watermarks are known to be quite vulnerable to pitch shifting attacks \cite{hu2014variable}.  Echo hiding is no exception; a shift in pitch up by a factor of $f$ will shift the echo down by a factor of $f$; therefore, we would expect degraded results in the presence of pitch shifting augmentation.  To quantify this, we design an experiment training RAVE on the Guitarset data embedded with a single echo at 75 samples, for varying degrees of pitch augmentation, and we test on the MUSDB18-HQ dataset as before.  Pitch shifting is disabled by default in RAVE, but when it is enabled, it randomly pitch shifts a clip 50\% of the time with simple spline interpolation at the sample level.  We modify the RAVE code to use higher quality pitch shifting with the Rubberband Library \cite{cannam2024pyrubberband}, and we enable a variable probability for pitch shifting.  When pitch shifting happens for a clip in a batch, we pick a factor uniformly at random in the interval $[0.75, 1.25]$.  Figure~\ref{fig:pitchshiftaugmentation} shows z-scores for training RAVE with an increasing probability of pitch shift augmentation, along with AUROC scores using the clean model to generate the false positive distribution.  As expected, the results degrade with increasing amounts of pitch shifting, though for the default value of 50\% pitch shifting, the z-scores are still quite far from the clean distribution.  Surprisingly, even at 90\% pitch shifting, the z-scores are still significant.

\subsection{Tagging Datasets}

\begin{figure}
    \centering
    \includegraphics[width=0.8\columnwidth]{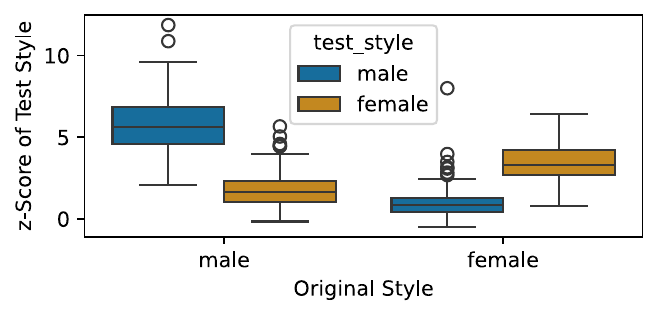}
    \caption{Tagging VocalSet, training with RAVE}
    \label{fig:vocalsettagging}
\end{figure}

We perform a preliminary experiment tagging a dataset with two different echoes depending on timbre: we tag all but one of the males in VocalSet with a 50 echo and all but one of the females in the dataset with a 75 echo.  As Figure~\ref{fig:vocalsettagging} shows, when we test with the remaining male and female, the z-scores of the corresponding echoes are higher.

\section{Discussion}
Overall, we have shown that an incredibly simple technique can be used to watermark training data; our implementations of single echo hiding and time spread echo hiding are each two lines of code in numpy/scipy.  One caveat is that, across all experiments, echoes are embedded more strongly in DDSP than in Rave, and in Rave than in Dance Diffusion, suggesting that complexity of the networks hampers the ability for the echoes to survive as strongly.  Still, each model reproduces the echoes to some degree, suggesting the generality of the approach.  This is surprising given how complex the models are and how they are unlikely to produce long sequences from the training data.

In future work, we would like to fine tune larger foundation models such as stable audio \cite{evans2024long} and to explore the extent to which different time spread echoes can simultaneously exist in different parts of such models.

\section{Acknowledgements}
We thank Bill Mongan and Leslie New for lending us computing resources for this project.  We also thank Tom Carroll via NSF-PHY-2011583 for compute cluster resources.

\bibliographystyle{plain}
\bibliography{writeup}

\end{document}